\documentclass[aps,final,notitlepage,oneside,twocolumn,nobibnotes,nofootinbib,
superscriptaddress,noshowpacs,centertags]{revtex4-1}

\usepackage[english]{babel}
\usepackage{graphicx}
\usepackage{latexsym}
\usepackage{amssymb}
\usepackage{amsmath}
\usepackage{float}

\begin{document}

\title{Destruction of Axion Miniclusters in the Galaxy}
\author{V. I. Dokuchaev}\thanks{e-mail: dokuchaev@inr.ac.ru}
\affiliation{Institute for Nuclear Research, Russian Academy of Sciences,
pr. 60-letiya Oktyabrya 7a, Moscow, 117312 Russia}
\affiliation{MEPhI National Research Nuclear University, Kashirskoe sh. 31, Moscow, 115409 Russia}
\author{Yu. N. Eroshenko}\thanks{e-mail: eroshenko@inr.ac.ru}
\affiliation{Institute for Nuclear Research, Russian Academy of Sciences,
pr. 60-letiya Oktyabrya 7a, Moscow, 117312 Russia}
\author{I. I. Tkachev}\thanks{e-mail: itkachev12@gmail.com}
\affiliation{Institute for Nuclear Research, Russian Academy of Sciences,
pr. 60-letiya Oktyabrya 7a, Moscow, 117312 Russia}
\affiliation{Novosibirsk State University, ul. Pirogova 2, Novosibirsk, 630090 Russia}

\date{\today}

\begin{abstract}
Previously, it has been established that axion dark matter (DM) is clustered to form clumps (axion
miniclusters) with masses $M\sim10^{-12}M_\odot$. The passages of such clumps through the Earth are very rare events
occurring once in $10^5$ years. It has also been shown that the Earth's passage through DM streams, which are
the remnants of clumps destroyed by tidal gravitational forces from Galactic stars, is a much more probable
event occurring once in several years. In this paper we have performed details calculations of the destruction
of miniclusters by taking into account their distribution in orbits in the Galactic halo. We have investigated
two DM halo models, the Navarro-Frenk-White and isothermal density profiles. Apart from the Galactic
disk stars, we have also taken into account the halo and bulge stars. We show that about 2-5\% of the axion
miniclusters are destroyed when passing near stars and transform into axion streams, while the clump
destruction efficiency depends on the DM halo model. The expected detection rate of streams with an
overdensity exceeding an order of magnitude is 1-2 in 20 years. The possibility of detecting streams by their
tidal gravitational effect on gravitational-wave interferometers is also considered.
\end{abstract}

\maketitle 

%\tableofcontents

%%%%%%%%%%%%

\section{INTRODUCTION}

Although dark matter (DM) accounts for $\simeq27$\% of
the mass of the Universe, its nature still remains
unknown. As yet undetected elementary particles are
considered as a probable candidate, and a number of
specific candidate particles, such as, for example, neutralinos,
sterile neutrinos, or axions, have been proposed
in theoretical works. The axion field was initially
introduced in particle physics to explain the
absence of CP violation in strong interactions. The
quanta of this field, axions, turned out to be promising
candidates for DM particles. Although the axions are
expected to have small masses, they belong to the type
of cold DM, because their production mechanism is
nonthermal; they were not in chemical and kinetic
equilibrium with the cosmic plasma.

The existence of DM clumps consisting of axions,
axion miniclusters, was predicted in \cite{KolTka94}. The clumps
are formed due to strong fluctuations of the axion field
in various regions of space on the horizon scale at the
epoch when the axion oscillations began. The fraction
of DM in the form of axion miniclusters is $f_{\rm mc}\sim1$.

A new effect that could increase the chances of
detecting the axion DM passing through the Earth was
presented in \cite{TinTkaZio16}. It was shown in \cite{TinTkaZio16} that although the
passage of a whole clump is an extremely rare event,
some of the clumps in the Galactic halo are destroyed
when interacting with stars, and DM streams with a
large overdensity from the destroyed clumps can be
observed in ground-based detectors approximately
once in 20 years. However, a simplified calculation
was performed in \cite{TinTkaZio16}. In particular, the orbital motion
of clumps and their distribution in orbits were disregarded.

The orbits of clumps in the Galactic halo are not
circular but, as a rule, eccentric. The clumps have
some distribution in their orbital parameters, with
their orbits undergoing precession (see \cite{BerDokEro07}, \cite{BerDokEro08}). If a
clump or its stream is now passing through the Solar
system, then it could previously pass closer to the
Galactic center, where the number density of stars is
larger and the destruction probability is higher. Therefore,
it is necessary to consider the passages through
the Galactic disk not only in the solar neighbourhood,
as was done in \cite{TinTkaZio16}, but also at other distances throughout
the entire life history of the clump in the Galactic
halo. The goal of this paper is to perform such a calculation.
The calculation technique used here is similar
to that applied in \cite{BerDokEro06}, \cite{BerDokEro07}, \cite{BerDokEro08}. In this paper we also take into
account the halo and bulge stars, which contribute
noticeably to the destruction of clumps.

The DM density and velocity in the Galactic halo
in the solar neighbourhood are largely fixed by the
observational data on the distribution and motion of
stars, because DM and baryons move in the same
gravitational potential. Nevertheless, there exists some
freedom in the choice of halo parameters, and the
observational data are compatible with various DM
halo models. To ascertain the dependence of the final
results on the halo model, we will perform calculations
for two distributions of DM particles and DM clumps
in the halo in their orbital parameters: the Navarro-Frenk-White and isothermal density profiles. The
Galactic halo model turned out to affect noticeably
the result.

In this paper we use the characteristic parameters
of clumps from \cite{KolTka94} in our calculations: the clump mass
$M=10^{-12}M_\odot$, the mean density $\bar\rho=140\rho_{\rm eq}\Phi^3(1+\Phi)$,
where $\Phi$ are the initial entropy density perturbations in
the medium of axions (for more details, see \cite{KolTka94}), at
$\Phi=1$ the clump radius is then $R=2.3\times10^{12}$~cm, while
the DM particle velocity dispersion in the clump is
$v_{\rm mc}\simeq(GM/R)^{1/2}\simeq7.6$~cm~s$^{-1}$ ($v_{\rm mc}/c\simeq2.5\times10^{-10}$). We
will choose the clump density profile in accordance
with the theory of multistream instability \cite{ufn1}:
\begin{equation}
 \rho_{\rm int}(r)=
 \frac{3-\beta}{3}\,\bar\rho\left(\frac{r}{R}\right)^{-\beta}.
 \label{rho}
\end{equation}
where $\beta=1.7-1.9$ (below in the calculations we set $\beta=1.8$).

\section{DESTRUCTION OF CLUMPS AND ADIABATIC PROTECTION}

A DM clump is destroyed if the net change in its
internal energy $\sum(\Delta
E)_j$) after one or more gravitational
interactions with disk stars or field is comparable
to the binding energy of the clump $|E|$:
\begin{equation}
 \sum\limits_j(\Delta E)_j\sim|E|,
 \label{prevcrit}
\end{equation}
where the summation is over the successive gravitational
interactions. In reality, the destruction occurs
not at once but there is a gradual mass loss predominantly
through the tidal stripping of outer DM layers
\cite{gnedin2,TayBab01,DieKuhMad,ZTSH,GoeGneMooDieSta07}. In this case, the central dense cores of the
clumps can survive \cite{BerDokEro08}.

If the internal revolution frequencies of DM particles
in their orbits inside the clump $\omega = v_{\rm mc}/R$ are
much higher than the characteristic frequency of the
external tidal force $\tau^{-1}$ (the reciprocal of the passage
time through the disk or the reciprocal of the passage
time of the impact parameter l to the star), then the
influence of the tidal force weakens significantly. This
effect is characterized by the so-called adiabatic correction
or the Weinberg correction $A(a)$, where $a=\omega\tau$, that is defined as the ratio of the energy change in
the real case to the energy change calculated in the
impulse approximation \cite{Wein1}. The following approximate formula was found in \cite{gnedin2}:
\begin{equation}
 A(a)=(1+a^2)^{-3/2}.
 \label{acor}
\end{equation}

The Galactic disk consists of stars and gas. The
destruction of clumps includes both the interaction
with the collective disk field and the interaction with
individual stars that happen to be near the clump trajectory
in the halo. In the former case, the tidal gravitational
field is produced not only by the disk stars but
also by the gas-dust clouds in the disk, i.e., by the total
gravitating disk mass.

Let us first consider the interaction of clumps with
the collective gravitational field of the disk when $\tau\simeq H_d/v$, where $H_d\sim 500$~pc is the Galactic disk halfthickness
and $v\sim200$~km~s$^{-1}$ is the characteristic passage
velocity. Whereas for neutralino clumps $a\sim1$ \cite{BerDokEro08},
for axion clumps $a\sim200$. Thus, in the latter case, the
clump destruction effect is suppressed approximately
by three orders of magnitude. The estimate made in
the impulse approximation shows that the destruction
time of axion miniclusters by the collective disk field is
$\sim10^{15}$~years. If the adiabatic correction is taken into
account, then this time increases to $\sim10^{18}$~years. Thus,
this destruction channel is inefficient, and during the
passage of axion clumps through the disk only the
interactions with individual stars are important, while
the gravitational shocks by the collective disk field play
no role.

During the interactions with individual stars $\tau\sim l/v$. Let us take the maximum impact parameter at
which the clump is destroyed in a single star flyby \cite{BerDokEro06}
as $l$:
\begin{equation}
 \left(\frac{l_*}{R}\right)^4=
 \frac{4(5-2\beta)}{3(5-\beta)} \frac{Gm_*^2}{MRv_{\rm rel}^2},
 \label{eqlzv}
\end{equation}
where $v_{\rm rel}$ is the relative velocity of the clump and the
star, and $m_*$ is the mass of the star. For the typical
parameters of clumps given at the end of the Introduction
and at $\beta\simeq2$, $v_{\rm rel}$=200~km~s$^{-1}$, $m_*\sim M_\odot$ we
obtain $l_*\sim500R$ and $a\sim10^{-4}$, i.e., in typical cases,
during the interactions with stars the adiabatic correction
is unimportant and will be disregarded below.

\section{DESTRUCTION OF CLUMPS IN GRAVITATIONAL COLLISIONS WITH STARS}

In \cite{BerDokEro06} the characteristic destruction time of a
clump as it moves in a medium of stars with number
density $n_*$ and mass $m_*$ was found for the case of
$l_*/R>1$:
\begin{equation}
 t_{*}=\frac{|E|}{\dot E}=
 \frac{1}{4\pi n_*m_*}
 \left[\frac{3(5-\beta)}{(5-2\beta)}\frac{M}{GR^3}\right]^{1/2}.
  \label{td1}
\end{equation}
Note that at $l_*/R>1$ the destruction time (\ref{td1}) does not
depend on the relative velocity of the clump and the
stars.

The survival probability of some specific clump
\begin{equation}
P_1=e^{-\int dt/t_*}
\label{p1eq}
\end{equation}
depends on its trajectory; therefore, we will first determine
the parameters of the clump trajectories in the
Galactic halo needed for the subsequent discussion in
a general form.

\subsection{The Trajectories of Clumps in the Galactic Halo}

Let us denote the orbital angular momentum of a
clump by $J$. The equation of the trajectory for $r(t)$ is
then
\begin{equation}
 M\dot r^2={2}\left[E_{\rm orb}-U(r)\right]-\frac{J^2}{Mr^2},
 \label{orbrazm}
\end{equation}
where $U(r)$ is the potential energy of the clump in the
halo. In what follows, we will consider an isotropic
distribution of clump orbits in the halo with a total
orbital energy $E_{\rm orb}$ (not to be confused with the internal
energy $E$) distribution function $F(E_{\rm orb})$ when the
relation to the halo density $\rho_{\rm H}(r)$ is given by the expressions
\cite{Edd16}
\begin{eqnarray}
 \rho_{\rm H}(r)&=&
 2^{5/2}\pi\int\limits^0_{U(r)}\sqrt{E_{\rm orb}-U(r)}F(E_{\rm orb})dE_{\rm orb}, \label{rhofe} \\
 F(E_{\rm orb})&=&\frac{1}{2^{3/2}\pi^2}\frac{d}{dE_{\rm orb}}
{  \int\limits_{r(E_{\rm orb})}^{\infty} }  \frac{dr}{\sqrt{E_{\rm orb}-U(r)}}
 \frac{d\rho_{\rm H}(r)}{dr},\;
 \label{ferho}
\end{eqnarray}
where the function $r=r(E_{\rm orb})$ is found from the equation
$U[r(E_{\rm orb})]=E_{\rm orb}$.

For the convenience of our subsequent discussion,
let us introduce the following dimensionless variables:
\begin{equation}
\xi=\frac{r}{R_c},  \quad  \tilde\rho(\xi)=\frac{\rho_{\rm H}^{~}(r)}{\rho_0},
\quad y=\frac{J^2}{2MU_0R_c^2},
\end{equation}
\begin{equation}
\varepsilon=\frac{E_{\rm orb}}{U_0}, \quad
\psi=\frac{U}{U_0},
\label{DimLessUnits}
\end{equation}
where $U_0$, $\rho_0$, and $R_c$ are some characteristic values of
the gravitational potential, the density, and the radius
in a specific DM halo model. In what follows, we will
choose these parameters so that $U_0=4\pi G\rho_0R_c^2M$.

The equation of the clump trajectory for the azimuthal
angle $\phi(\xi)$ is
\begin{equation}
 \frac{d\phi}{d\xi}=
 \frac{y^{1/2}}{\xi^2\sqrt{\varepsilon-\psi(\xi)-y/\xi^2}}.
\end{equation}
The equation for the extreme points of the orbit $\dot r^2=0$
will be written as
\begin{equation}
 \frac{y}{\xi^2}=\varepsilon-\psi.
 \label{minmaxbezr}
\end{equation}
In our calculations we then numerically find the roots
of this equation $\xi_{\rm min}$ and $\xi_{\rm max}$. Twice the time of clump
motion from $\xi_{\rm min}$ to $\xi_{\rm max}$
\begin{equation}
 T_c(\varepsilon,y)=\frac{1}{\sqrt{2\pi G\rho_0}}
 \int\limits_{\xi_{\rm min}}^{\xi_{\rm max}}
 \frac{d\xi}{\sqrt{\varepsilon-\psi(\xi)-y/\xi^2}}
 \label{pcintnfw}
\end{equation}
is not equal to the orbital period, because the orbital
precession should be additionally taken into account.
The precession angle in the time $T_{\rm c}/2$ is
\begin{equation}
 \tilde\phi=y^{1/2}\int\limits_{\xi_{\rm min}}^{\xi_{\rm max}}
 \frac{d\xi}{\xi^2\sqrt{\varepsilon-\psi(\xi)-y/\xi^2}}-\pi,
 \label{precnfw}
\end{equation}
and $\tilde\phi<0$. Therefore, the true period (the revolution
around the Galactic center through $2\pi$) is
\begin{equation}
 T_{\rm t}=T_{\rm c}\left(1+\tilde\phi/\pi\right)^{-1}.
 \label{pcinttnfw}
\end{equation}

Below we consider the clumps whose orbits are
currently passing through the Solar system at a distance
$r=r_\odot=8.5$~kpc from the Galactic center.
Denote $p=\cos\theta$, where $\theta$ is the angle between the
radius vector of the clump $\vec r$ and its velocity $\vec v$ in the
solar neighborhood. The dimensionless parameter $y$
characterizing the angular momentum of the clump
can then be found from the expression
\begin{equation}
 y=(1-p^2)\xi^2\left[\varepsilon-\psi(\xi)\right],
 \label{yexpnfw}
\end{equation}
where we should set $\xi=r_\odot/R_c$.

\subsection{Destruction of Clumps by Disk Stars}
\label{nfwdisksubsec}

In the lifetime of the Galaxy $t_{\rm G}\simeq10^{10}$~years a clump
experiences $N\simeq t_{\rm G}/T_{\rm t}$ double crossings of the Galactic
disk, with the crossing points every time being shifted
by angles $|\tilde\phi|$ due to the precession effect.

The survival probability (\ref{p1eq}) of some specific clump
contains the integral
\begin{equation}
\int m_*n_*~dt\simeq \sum \int \frac{m_* n_* dl}{v},
\label{sumnt}
\end{equation}
where the summation is over the successive Galactic
disk crossings in the time $t_{\rm G}$, while the integration is
over one specific crossing. This integral is expressed
via the mass surface density $\sigma_s$ of the stellar component
of the Galactic disk,
\begin{equation}
\int \frac{m_* n_* dl}{v}=\frac{\sigma_s}{ v_z},
\end{equation}
while the distribution of stars in masses $m_*$ does not
enter into the result. The clump velocity component
along the normal to the disk is written as
\begin{equation}
 v_z=\frac{J}{r}\sin\gamma,
 \label{vzc}
\end{equation}
where $\gamma$ is the angle between the normal to the orbital
plane and the normal to the Galactic disk plane. The
surface density of the stellar component of the Galactic
disk at point $r$ of its crossing by the clump is given
by the expression
\begin{equation}
  \label{diskmass}
\sigma_s(r)=\frac{M_d}{2\pi r_0^2}\,e^{-r/r_0},
\end{equation}
where $M_d=3\times10^{10}M_{\odot}$ and $r_0=4.5$~kpc, so that
$\sigma_s(r_\odot)=35M_{\odot}$~pc$^{-2}$. Here, we take into account the
fact that the stars constitute only part of the total disk
mass. The normalization $\sigma_s(r_\odot)=35M_{\odot}$~pc$^{-2}$ corresponding
to the stars was taken from \cite{KuiGil89} (page 635).

The orbital precession effect facilitates considerably
the calculation of the sum in Eq.~(\ref{sumnt}), because the
clump successively crosses the disk at equal angular
intervals at all radii between the minimum and maximum
radial distances of the orbit owing to the precession.
Therefore, we approximately calculate the sum
as follows:
\begin{equation}
 \sum\limits_{i=1}^{N}\sigma_s(r)r\simeq
\frac{1}{|\tilde\phi|}\int \sigma_sR_c\xi d\phi
 \simeq\frac{R_c}{|\tilde\phi|}\int\limits_{\xi_{\rm min}}^{\xi_{\rm max}}
 \sigma_s(\xi)\xi\frac{d\phi}{d\xi}d\xi\frac{2t_{\rm G}}{T_{\rm t}},
 \nonumber
\end{equation}

Let us rewrite (\ref{p1eq}) as
\begin{equation}
P_1=\exp\left\{-\frac{\lambda}{\Phi^{3/2}(1+\Phi)^{1/2}\sin\gamma}\right\},
 \label{a1disknfw}
\end{equation}
where
\begin{equation}
\lambda=2(2\pi)^{1/2}\left(\frac{5-2\beta}{5-\beta}\right)^{1/2}\frac{G^{1/2}t_{\rm G}S}{T_{\rm t}|\tilde\phi|(140\rho_{\rm eq})^{1/2}U_0^{1/2}},
\end{equation}
\begin{equation}
S=\int\limits_{\xi_{\rm min}}^{\xi_{\rm max}}
 \frac{d\xi\sigma_s(\xi)}{\xi\sqrt{\varepsilon-\psi(\xi)-y/\xi^2}}.
\end{equation}
Using (\ref{pcintnfw}), we obtain the fraction of destroyed clumps
in the solar neighbourhood:
\begin{equation}
P =1- 
\frac{ 
 \int\limits_{0}^{1}dp\int\limits_{0}^{\sin\alpha}d\cos\gamma
 \int\limits_{\psi(\xi)}^{0}d\varepsilon
 [\varepsilon-\psi(\xi)]^{1/2}F(\varepsilon)P_1,
}{ 
 \int\limits_{0}^{1}dp\int\limits_{0}^{\sin\alpha}d\cos\gamma
 \int\limits_{\psi(\xi)}^{0}d\varepsilon
 [\varepsilon-\psi(\xi)]^{1/2}F(\varepsilon),
} 
 \label{pd1nfw}
\end{equation}
where we should substitute $\xi=r_\odot/R_c$ and $\alpha\approx\pi/2$.

%%%%%%%%%%%%

   \subsection{Destruction of Clumps by Halo and Bulge Stars}
   \label{nfwstarsubsec}   
   
Outside the Galactic disk there are stars of the
spherical subsystems: these are halo and bulge stars
(plus stars in globular clusters, which we disregard).
The number density of stars in the halo at a distance
$r>3$~kpc from the Galactic center is
\begin{equation}
  n_{h,*}(r)=(\rho_\odot/m_*) (r_{\odot}/r)^{3},
 \label{rhosh}
\end{equation}
where we took $\rho_{\odot}=10^{-4}~M_{\odot}/$pc$^3$ as an estimate. Note,
however, that an order of magnitude larger value, $\rho_{\odot}=10^{-3}~M_{\odot}/$pc$^3$, is obtained in some studies (see \cite{ColOst81} and
references in \cite{ColOst81}, \cite{MarSuch}). However, the authors of \cite{MDSQ2}
point out that even $\rho_{\odot}=10^{-4}~M_{\odot}/$pc$^3$ should be considered
as an upper limit for the density of stars in the
halo.

The number density of stars in the bulge at a distance
$r=1-3$~kpc \cite{LauZylMez} is
\begin{equation}
 n_{b,*}(r)=(\rho_b/m_*)\exp\left[ -(r/r_b)^{1.6}\right],
 \label{rhoe}
\end{equation}
where ?$\rho_b=8M_{\odot}/$pc$^3$ and $r_b=1$~kpc.

For our calculations we will need to sum the clump
energy change over the orbital period or, which is
mathematically equivalent, to average $t^{-1}_*$ over the
clump trajectory in the halo:
\begin{equation}
\langle t_*^{-1}\rangle=\frac{R_c}{ T_c} \sqrt{\frac{2}{U_0}}
 \int\limits_{\xi_{\rm min}}^{\xi_{\rm max}}
 \frac{t_*^{-1}d\xi}{\sqrt{\varepsilon-\psi(\xi)-y/\xi^2}}.
 \label{tzvintnfw}
\end{equation}
The survival probability of a single clump is
\begin{equation}
P_1=e^{-t_{\rm G}\langle t_*^{-1}\rangle}=\exp\left\{-\frac{\lambda}{\Phi^{3/2}(1+\Phi)^{1/2}}\right\},
 \label{nfwstarp1}
\end{equation}
where, in the case under consideration,
\begin{equation}
\lambda=2(2\pi)^{1/2}\left(\frac{5-2\beta}{5-\beta}\right)^{1/2}\frac{R_c m_*G^{1/2}t_{\rm G}S}{T_{\rm c}(140\rho_{\rm eq})^{1/2}U_0^{1/2}},
\end{equation}
\begin{equation}
 S=\int\limits_{\xi_{\rm min}}^{\xi_{\rm max}}
 \frac{d\xi n_s(\xi)}{\sqrt{\varepsilon-\psi(\xi)-y/\xi^2}}.
\end{equation}
Owing to the presumed spherical symmetry of the halo
and the bulge, the expression for the fraction of
destroyed clumps is simplified:
\begin{equation}
P =1- 
\frac{ 
 \int\limits_{0}^{1}dp
 \int\limits_{\psi(\xi)}^{0}d\varepsilon
 [\varepsilon-\psi(\xi)]^{1/2}F(\varepsilon)P_1
}{ 
 \int\limits_{0}^{1}dp
 \int\limits_{\psi(\xi)}^{0}d\varepsilon
 [\varepsilon-\psi(\xi)]^{1/2}F(\varepsilon)
}. 
 \label{pdstarnfw}
\end{equation}

%%%%%%%%%%%%

\section{DESTRUCTION OF CLUMPS IN THE NAVARRO-FRENK-WHITE HALO MODEL}

Let us first calculate the destruction of axion miniclusters
for the Navarro-Frenk-White density profile
\begin{equation}
 \rho_{\rm H}(r)=\frac{\rho_0}{(r/R_c)(1+r/R_c)^2},
 \label{nfwhalo}
\end{equation}
where $\rho_{\rm H}(r_\odot)=0.3$~GeV/cm$^3$ and $R_c=20$~kpc. It
should be noted that a density profile close in shape to
the Navarro-Frenk-White profile was obtained in the
analytical model \cite{SikTkaWan96}.

The halo density in dimensionless variables is
\begin{equation}
\tilde\rho(\xi)=\frac{1}{\xi(1+\xi)^2}.
 \label{nfwhalox}
\end{equation}
Choosing $U_0=4\pi G\rho_0R_c^2$, we find the gravitational
potential in dimensionless variables:
\begin{equation}
 \psi(\xi)= -\frac{\log(1+\xi)}{\xi}.
 \label{potpsi}
\end{equation}

The distribution function $F(\varepsilon)$ for the profile (\ref{nfwhalox})
was approximated in \cite{Wid00} by the expression
\begin{equation}
F(\varepsilon)=F_1(-\varepsilon)^{3/2}(1+\varepsilon)^{-5/2}
\left[-\frac{\ln(-\varepsilon)}{(1+\varepsilon)}\right]^qe^P,
 \label{fffrhofd}
\end{equation}
where $F_1=9.1968\times10^{-2}$,
$P=\sum\limits_{i}p_i(-\varepsilon)^i$,
$(q{,}p_1{,}p_2{,}p_3{,}p_4)=(-2.7419{,} 0.3620{,} -0.5639{,}-0.0859{,} -0.4912)$.
Then,
\begin{equation}
\tilde\rho(\xi)=4\pi\sqrt{2}\int\limits_{\psi(\xi)}^{0}d\varepsilon
\left[\varepsilon-\psi(\xi)\right]^{1/2}F(\varepsilon).
 \label{rhofd}
\end{equation}

\begin{figure}%[h]
	\begin{center}
\includegraphics[angle=0,width=0.49\textwidth]{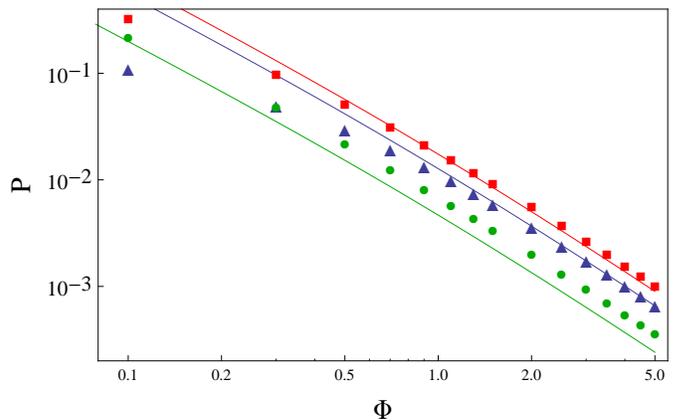}
	\end{center}
\caption{Fraction of destroyed axion miniclusters during
their interactions with disk stars versus density perturbation
$\Phi$ for the Galactic halo with the Navarro-Frenk-White density profile. The circles indicate the result of our
calculation using Eq.~(\ref{pd1nfw}), while the solid line indicates
the approximate expression (\ref{pd2nfw}). The triangles indicate the
result of our exact calculation using Eq.~(\ref{pdstarnfw}), while the
corresponding solid line indicates the approximate expression
(\ref{starnfwupr}). The total fraction of destroyed axion miniclusters
is indicated by the squares, while the solid line passing
through them indicates the sum of (\ref{pd2nfw}) and (\ref{starnfwupr}).}
	\label{grsnfw}
\end{figure}

The fraction of clumps in the solar neighbourhood
destroyed in their collisions with stars (\ref {pd1nfw}) found by
numerically calculating all of the integrals in it is indicated
by the circles in Fig.~\ref{grsnfw} for various values of $\Phi$. If
the quantity in the exponent in (\ref{a1disknfw}) is much smaller
than unity in absolute value, then we can expand the
exponential into a series $e^x\approx1+x$ and take the integral
(\ref{pd1nfw}) over $\gamma$ analytically and the remaining integrals
numerically. This allows the functional dependence
on $\Phi$ to be separated out. The result of such a calculation
is
\begin{equation}
P = \frac{6.6\times10^{-3}}{\Phi^{3/2}(1+\Phi)^{1/2}}
 \label{pd2nfw}
\end{equation}
and is indicated in Fig.~\ref{grsnfw} by the solid line. It can be
seen that, in this case, there is some difference
between the exact and approximate expressions.

Our calculation of the destruction by halo and
bulge stars using Eq.~(\ref{pdstarnfw}) is indicated by the triangles
in Fig.~\ref{grsnfw}. If the exponential in (\ref{nfwstarp1}) can be expanded,
then, as above, we approximately obtain
\begin{equation}
P^{(s)}=\frac{1.8\times10^{-2}}{\Phi^{3/2}(1+\Phi)^{1/2}}.
 \label{starnfwupr}
\end{equation}
This quantity is indicated in Fig.~\ref{grsnfw} by the solid line. It
can be seen that at $\Phi\geq1$ the quantity (\ref{starnfwupr}) serves as a
good approximation to the exact numerical result.

The total fraction of destroyed axion miniclusters,
including their destructions by disk, halo, and bulge
stars, is indicated by the squares in Fig.~\ref{grsnfw} and is satisfactorily
described by the sum of Eqs.~(\ref{pd2nfw}) and (\ref{starnfwupr}).

Comparing (\ref{pd2nfw}) with Eqs.~(3.3) from \cite{TinTkaZio16}, we see
that the numerical calculation performed in this paper
gives approximately a factor of 3 smaller fraction of
destroyed clumps if only the destruction by Galactic
disk stars is taken into account. The difference
between the results is explained by the fact that, in
reality, the orbits of clumps in the halo are noncircular
and a predominant fraction of the clumps crossing the
orbit of the Solar system today spent most of the time
at a distance from the Galactic center larger than the
distance from the center to the Sun (as was suggested
in \cite{TinTkaZio16}). The disk crossings in the outer region of the
Galaxy, where the disk has a lower surface density,
exert a smaller destructive effect on the clumps. However,
additional destruction is caused by halo and
bulge stars, which, as a result (in the sum with (\ref{starnfwupr})),
leads to an increase in the total fraction of destroyed clumps by 25\%. Thus, the final result turns out to be
close to the result of \cite{TinTkaZio16}, where only the Galactic disk
stars were taken into account and no halo and bulge
stars were considered.

\section{THE ISOTHERMAL DENSITY PROFILE}

To ascertain how the result obtained depends on
the Galactic halo model, let us perform calculations
similar to the previous ones, but for the isothermal
density profile of the Galactic halo
\begin{equation}
 \rho_{\rm H}(r)=\frac{1}{4\pi}\frac{v_{\rm rot}^2}{Gr^2},
 \label{pureiso}
\end{equation}
where $v_{\rm rot}=(GM_{\rm H}/R_{\rm H})^{1/2}$, $R_{\rm H}\simeq200$~kpc, and $\rho(r)=0$
at $r>R_{\rm H}$. We choose $R_c = R_{\rm H}$; in this case, $U_0 = v_{\rm rot}^2$
and the potential in dimensionless variables (\ref{DimLessUnits}) is
\begin{equation}
 \psi(r)=\log(\xi).
 \label{pot}
\end{equation}

Using (\ref{ferho}) for the profile (\ref{pureiso}) with the boundary at
$r=R_{\rm H}$, we obtain
\begin{equation}
 F(\varepsilon)=
 \frac{1}{2^{5/2}\pi^3e}
 \frac{v_{\rm rot}^{1/2}}{GM^{3/2}R_{\rm H}^2}\,F(\varepsilon)
 \label{feerf},
\end{equation}
where
\begin{equation}
 f(\varepsilon)=\sqrt{2\pi}\,e^{-2\varepsilon+2}\,{\rm erf}\left[\sqrt{-2\varepsilon}\right]\!+\!
 \frac{e^2}{\sqrt{-\varepsilon}}
 \label{jerf}.
\end{equation}
Note that this distribution function does not reproduce
the isothermal profile exactly.

\begin{figure}%[h]
	\begin{center}
\includegraphics[angle=0,width=0.49\textwidth]{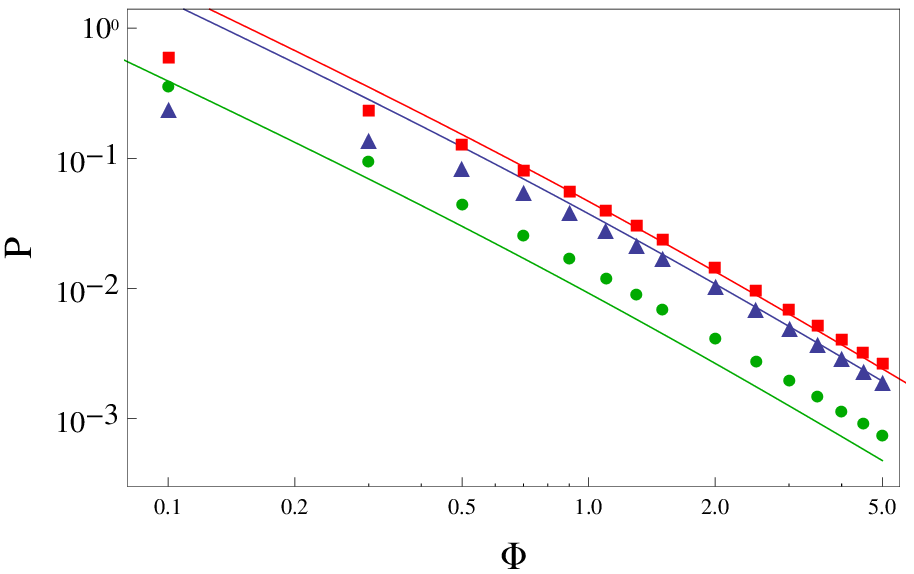}
	\end{center}
\caption{Fraction of destroyed axion miniclusters during
their interactions with disk stars versus density perturbation
$\Phi$ for the Galactic halo with the isothermal density
profile. The circles indicate the result of our calculation
using Eq.~(\ref{pd1nfw}), while the solid line indicates the approximate
expression (\ref{pd2}). The triangles indicate the result of
our exact calculation using Eq.~(\ref{pdstarnfw}), while the corresponding
solid line indicates the approximate expression (\ref{pdstariso}).
The total fraction of destroyed axion miniclusters is indicated
by the squares, while the solid line passing through
them indicates the sum of (\ref{pd2}) and (\ref{pdstariso}).}
	\label{grsiso}
\end{figure}

\begin{figure}%[h]
	\begin{center}
\includegraphics[angle=0,width=0.49\textwidth]{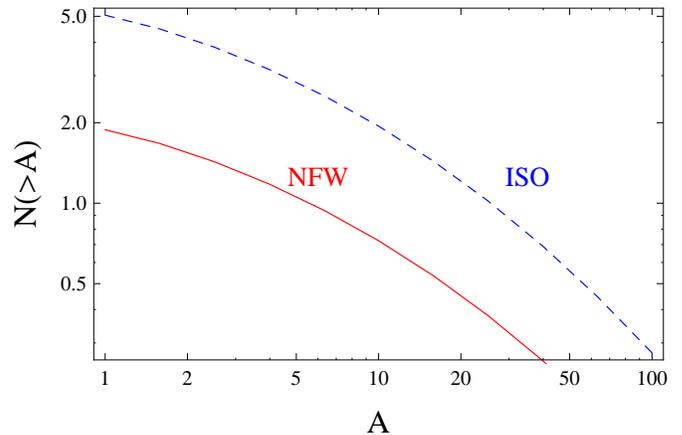}
	\end{center}
\caption{Detection rate of axion bursts (the number of bursts
detected in 20 years of observations) with a density amplification
larger than $A$. The solid curve indicates the result
of our calculation in the case of the Navarro-Frenk-White profile for the sum of (\ref{pd2nfw}) and (\ref{starnfwupr}). The dashed
curve indicates the result of our calculation in the case of
the isothermal halo density profile for the sum of (\ref{pd2})
and (\ref{pdstariso}).}
	\label{grrate}
\end{figure}

The fraction of clumps in the solar neighbourhood
destroyed in their collisions with stars (\ref {pd1nfw}) found by
numerically calculating all of the integrals in it is indicated
by the dots in Fig.~\ref{grsiso} for various values of $\Phi$. If the
quantity in the exponent in (\ref{a1disknfw}) is much smaller than
unity in absolute value, then
\begin{equation}
P \simeq\frac{1.3\times10^{-2}}{\Phi^{3/2}(1+\Phi)^{1/2}}.
 \label{pd2}
\end{equation}
This result is indicated in Fig.~\ref{grsiso} by the solid line.

The results of our calculation of the destruction by
halo and bulge stars using Eq.~(\ref{pdstarnfw}) are indicated in
Fig.~\ref{grsiso} by the triangles. If the exponential in (\ref{nfwstarp1}) can be
expanded, then, as above, we approximately obtain
\begin{equation}
P^{(s)}=\frac{5.3\times10^{-2}}{\Phi^{3/2}(1+\Phi)^{1/2}}.
 \label{pdstariso}
\end{equation}
This quantity and the total quantities for the isothermal
density profile are shown in Fig.~\ref{grsiso}.

\section{OBSERVATIONAL CONSEQUENCES}

\subsection{Detection of Streams in Axion Detectors}

We calculate the expected detection rate of streams
in ground-based detectors in the same way as was
done in \cite{TinTkaZio16}. According to \cite{TinTkaZio16} (with the correction
coefficient $3/2$ in Eq.~(4.5) from \cite{TinTkaZio16}), the frequency of
stream-crossing events is
\begin{equation}
d\nu=\frac{3P_{\rm mc}(\Phi)[P(\Phi)+P^{(s)}(\Phi)]a(\Phi)}{2\tau(\Phi)A^3}dAd\Phi,
 \label{dnu}
\end{equation}
Here, $P_{\rm mc}(\Phi)$ is the distribution of axion miniclusters
in perturbations $\Phi$, $a(\Phi)$ is the overdensity in the ministream
with respect to the mean DM density in the
Galactic halo in the solar neighbourhood $\rho_{\rm H}(r_\odot)$ in the
case where the minicluster is destroyed immediately
after the Galactic disk formation (for more details, see
\cite{TinTkaZio16}), $A$ is the real overdensity in the ministream, and
$\tau(\Phi)=2R/v$ is the passage time of the Earth through
the ministream cross section. To obtain the detection
rate of bursts $N(>A)=\nu(A)\Delta t$ with a density amplification
larger than $A$ in an observation time $\Delta t$, it is necessary
to integrate (\ref{dnu}) over $\nu$ from ${\rm max}(A,a(\Phi))$ to
$\bar{\rho}(\Phi)/\rho_{\rm H}(r_\odot)$ and over all $\Phi$. The result of our calculations
is indicated in Fig. 3 by the lower and upper lines
for the Navarro-Frenk-White and isothermal density
profiles, respectively.

Thus, we see that there is a dependence of the
results on the Galactic halo model. For the Navarro-Frenk-White profile the destruction of clumps by the
disk is approximately half as efficient as that for the
singular isothermal halo. For the destruction by halo
stars the isothermal profiles gives an almost a factor of
3 larger value.

\subsection{On the Possibility of Detecting Streams by the LISA Detector}

If a stream passes through the Solar system, then its
gravitational field will act on gravitational-wave interferometers.
The relative length of the interferometer
arm $\Delta l/l$ will change under the tidal gravitational force
from the stream. It is interesting to consider such an
action on the planned LISA interferometer, which is
expected to have a very high sensitivity, $\sim2\times10^{-18}$.
The signals in the interferometer will be in the form of
single pulses. The pulse structure in three directions
will be strictly synchronized with the signals in the
ground-based axion detectors. Therefore, based on
the pattern of the pulses, it will be possible to prove
almost unambiguously the passage of a stream and to
ascertain its velocity direction and overall structure.
The possibility of detecting compact objects with
masses $10^{14}-10^{20}$~g using LISA was pointed out in \cite{LISAPBH}, \cite{LISAAsteroid},
\cite{LISADM}, where primordial black holes, asteroids, or massive
DM objects were considered as compact objects.
In contrast to these papers, in our case, it is necessary
to consider a noncompact mass distribution in the
form of an elongated stream.

We model the stream by a straight thin thread of
length $L=v_{\rm mc}t$, where $v_{\rm mc}$ is the internal velocity dispersion
in the clump and $t$ is the time elapsed since the
clump destruction. The gravitational field of the
stream at distance $r$ from the axis is then
\begin{equation}
g=\frac{2GM}{rL}.
\end{equation}
If $l\sim5\times10^{11}$~cm is the interferometer arm length
(in the new eLISA project the arm length was reduced
to $1\times10^{11}$~cm), then the tidal acceleration is
\begin{equation}
\Delta g\sim\frac{2GM}{r^2L}l,
\end{equation}
while the change of the arm length in the stream passage
time $\Delta t\sim r/v_{\rm rel}$ is
\begin{equation}
\Delta l\sim\Delta g (\Delta t)^2/2,
\end{equation}
where $v_{\rm rel}\sim200$~km~s$^{-1}$. The relative change of the arm
is
\begin{equation}
\frac{\Delta l}{l}\sim\frac{GM}{v_{\rm rel}^2v_{\rm mc}t}\sim3\times10^{-19}
\label{dll}
\end{equation}
at $t\sim5\times10^9$~years. The quantity (\ref{dll}) does not depend
on $r$ and is comparable to the LISA sensitivity. Slow
streams with a lower $v_{\rm rel}$ will act on the detector more
efficiently, but their number is also smaller. Axion
streams will produce additional ``noise'' in space-borne
detectors. The same expression for $\Delta l/l$ as (\ref{dll}) is also
obtained if the passage of the interferometer arm
inside a stream is considered.

To assess more accurately the prospects for the
detection of axion streams by gravitational-wave interferometers,
we will take into account the detector
noise distribution. In our calculation we follow the
method described in \cite{LISAPBH}. Let $r_{\rm min}$ be the minimum
distance from the axis of the passing stream to the center
of the segment connecting the two interferometer
mirrors. We will assume that during the passage the
detector is always outside the stream. The tidal gravitational
acceleration that the interferometer arm experiences
is
\begin{equation}
a(t)=\frac{2GMl}{L[r_{\rm min}^2+(tv_{\rm rel})^2]},
\end{equation}
while its Fourier spectrum is
\begin{equation}
a(f)=\int\limits_{-\infty}^{+\infty}dte^{2\pi ift}a(t)=\frac{2GMl}{Lv_{\rm rel}r_{\rm min}}e^{-2\pi r_{\rm min}f/v_{\rm rel}}.
\end{equation}
If the detection is based on the optimal filtering
method, then for the square of the signal-to-noise
ratio we have
\begin{equation}
\rho^2_{\rm SN}=4\int\limits_{0}^{+\infty}df\frac{a^2(f)}{S^2(f)},
\end{equation}
where for the LISA detector $S\simeq a_0\simeq6\times10^{-13}$~cm~s$^{-2}$~Hz$^{-1/2}$. Assuming for the estimate that
$S=a_0=const$, we obtain
\begin{equation}
\rho^2_{\rm SN}=\frac{4\pi G^2M^2l^2}{a_0^2L^2v_{\rm rel}r_{\rm min}^3}.
\end{equation}
To detect a stream with a given $\rho_{\rm SN}$, it is necessary that
the stream pass at a distance no larger than $r_{\rm min}$ from
the detector. We numerically obtain
\begin{eqnarray}
&&r_{\rm min}=6\times10^{12}\left(\frac{l}{5\times10^{12}\mbox{~cm}}\right)^{2/3} \nonumber \\
&\times&\left(\frac{a_0}{6\times10^{-13}\mbox{~cm~s$^{-2}$~Hz$^{-1/2}$}}\right)^{-2/3}\left(\frac{\Delta t}{5\times10^9\mbox{~yars}}\right)^{-2/3}
\nonumber 
\\
&\times&\left(\frac{v_{\rm rel}}{200\mbox{~km~s$^{-1}$}}\right)^{-2/3}\left(\frac{\rho_{\rm SN}}{0.05}\right)^{-2/3}~\mbox{cm},
\label{rmineq}
\end{eqnarray}
With these normalization values the detection rate of
streams will be
\begin{equation}
{\rm Rate}=\pi r_{\rm min}^2v_{\rm rel}\frac{f_{\rm mc}\rho_{\rm DM}P}{M}\frac{v_{\rm mc}\Delta t}{R}\sim0.1\mbox{~year$^{-1}$},
\label{rate}
\end{equation}
where$f_{\rm mc}\sim1$ is the fraction of DM in the form of axion
miniclusters and $P\sim0.02$ is the minicluster destruction
probability calculated in previous sections.

Let us first consider the case with a single LISA-type
detector. If we assume in (\ref{rmineq}) that the signal-to-noise
ratio is $\rho_{\rm SN}\sim5$, as is commonly assumed for a
single detector, and choose $l=5\times10^{11}$~cm, then (\ref{rmineq})
is smaller than the stream radius approximately by two
orders of magnitude. The detection rate would be
$\sim10^{-5}$~yr$^{-1}$. Thus, next-generation detectors, in which
the interferometer arm $l$ is larger than that in LISA by
one and a half or two orders of magnitude and the
noise $a_0$ is lower, are needed for the detection of
streams with an acceptable rate. Allowance for the
detector passage inside a stream and for the distribution
in $v_{\rm rel}$, probably, will not change greatly the result.
In the new eLISA project the detector noise at low frequencies
is very large \cite{Amaetal12}; at a characteristic frequency
of $10^{-5}$~Hz we have $a_0\simeq1.5\times10^{-11}$~cm~s$^{-2}$~Hz$^{-1/2}$. Therefore,
in comparison with LISA, the detection rate of
streams will be lower approximately by two more
orders of magnitude. The dependence of the result on
the distribution of clumps in velocities and directions
is also interesting in the problem of the detection of
clumps by gravitational-wave interferometers, but
these questions are beyond the scope of this paper.

However, we may consider the detection of streams
in two detectors, if there are two or more orbiting
interferometers of the next (after LISA) generation, by
the coincidence method and the characteristic signal
shape. Suppose that the interferometer arm is an order
of magnitude larger than was planned in LISA. In this
case, a variant with $\rho_{\rm SN}<1$ is admissible, which was
chosen in the normalization coefficients in (\ref{rmineq}) and
(\ref{rate}). In this case, one might expect an acceptable
detection rate from the viewpoint of real observations.

\section{CONCLUSIONS}

The particles being lost by a clump during its gravitational
interactions with stars form a stream behind
the clump being destroyed. In this way the bulk of the
mass or the entire mass of the clump can pass into the
stream. Since the area of the stream is larger than that
of the clump by several orders of magnitude, the
Earth's passage through the stream is a much more
probable event than its passage through the whole
clump. Therefore, allowance for the streams is of great
and, possibly, fundamental importance for the experiments
aimed at directly detecting axion DM particles,
as was shown in \cite{TinTkaZio16}.

In this paper we performed a calculation similar to
that in \cite{TinTkaZio16}, but with allowance made for two additional
effects. First, we took into account the fact that the
orbits of clumps in the halo are noncircular and precess
and that throughout its life history a clump could
cross the Galactic disk at different distances from the
center and could pass through halo regions with different
number densities of stars. We considered two halo
models, the Navarro-French-White profile and an
isothermal sphere, and showed that the destruction in
the second model is approximately a factor of two or
three more efficient. Thus, the halo model affects
noticeably the result. This influence is related to a different
distribution of DM clumps in their orbital
parameters.

The Navarro-Frenk-White profile was obtained
in the numerical simulations of galaxies without
including the baryonic component. The cooling of
baryons and their settling to the halo center must lead
to a deepening of the potential well and an additional
increase of the DM density in the central halo region.
Therefore, it is possible that the isothermal profile
corresponds better to the real one, because in it the
density concentration at the center is larger than that
in the Navarro-Frenk-White halo.

Second, we took into account the destruction of
clumps by Galactic halo and bulge stars. This effect
increases the overall destruction efficiency. As is easy
to show, the destruction of clumps during their pair
interactions with one another is several orders of magnitude
less efficient than that during the interactions
of clumps with halo stars.

As a result, we obtained the distribution of the rate
of stream-crossing events as a function of overdensity.
For example, we found that at an overdensity $A>10$
one might expect 1-2 events in 20 years.

The prospects for the detection of streams from
destroyed clumps with gravitational-wave interferometers
look realistic only for detectors of the next (after
LISA) generation or in the case of several LISA-type
detectors and using a detection technique based on the
coincidence method at a signal-to-noise ratio much
less than unity in one detector.

We are grateful to D. Levkov and A. Panin for the
useful discussions. This study was supported by the
Russian Science Foundation (project no. 16-12-10494).

\end{document}